\documentclass[12pt, letterpaper]{article}
\usepackage{graphicx}
\usepackage{tabularx}
\usepackage{url}

\title{Identification of Significant Permissions for Efficient Android Malware Detection}
\author{
Hemant Rathore, Sanjay K. Sahay, Ritvik Rajvanshi\\
Department of CS\&IS, BITS Pilani, Goa, India\\
\texttt{\{hemantr, ssahay, f20160544\}@goa.bits-pilani.ac.in}
\and
Mohit Sewak\\
Security \& Compliance Research, Microsoft R\&D, India\\
\texttt{mohit.sewak@microsoft.com}
}

\date{} 

\begin{document}
\maketitle

\begin{abstract}
Since Google unveiled Android OS for smartphones, malware are thriving with 3Vs, i.e. volume, velocity and variety. A recent report indicates that one out of every five business/industry mobile application leaks sensitive personal data. Traditional signature/heuristic based malware detection systems are unable to cope up with current malware challenges and thus threaten the Android ecosystem. Therefore recently researchers have started exploring machine learning and deep learning based malware detection systems. In this paper, we performed a comprehensive feature analysis to identify the significant Android permissions and propose an efficient Android malware detection system using machine learning and deep neural network. We constructed a set of $16$ permissions ($8\%$ of the total set) derived from variance threshold, auto-encoders, and principal component analysis to build a malware detection engine which consumes less train and test time without significant compromise on the model accuracy. Our experimental results show that the Android malware detection model based on the random forest classifier is most balanced and achieves the highest area under curve score of $97.7\%$, which is better than the current state-of-art systems. We also observed that deep neural networks attain comparable accuracy to the baseline results but with a massive computational penalty.

\end{abstract}
\section{Introduction}
Today smartphones have penetrated very deep into our modern society and touch the lives of more than $60\%$ of the world population \cite{globaldr}. There has been an explosive growth in new smartphones sold every year (from $173.5$ million in $2009$ to $1,474$ million in $2017$) \cite{statista_sold}. Smartphone uses an Operating System (OS) for resource management. Currently, Android OS from Google holds more than $80\%$ of market share, followed by iOS from Apple at $14.9\%$ \cite{globaldr}. The broad acceptance of Android OS is due to its open-ecosystem, large application base, multiple app stores, and customizability. Around $50\%$ of smartphone customers connect to the internet, which is currently the primary attack vector for malware developers \cite{ye2017survey} \cite{tam2017evolution}.

\textbf{Mal}icious Soft\textbf{ware} (a.k.a. Malware) are not new to the current digital world. The first malware was a worm named \textit{Creeper}\footnote{\url{https://www.trendmicro.com/vinfo/us/threat-encyclopedia/archive/malware/creeper.472.b}}, an experimental self-replicating program written for fun in $1971$ \cite{ye2017survey}. On mobile devices, \textit{Cabir}\footnote{\url{https://www.f-secure.com/v-descs/cabir\_dropper.shtml}} ($2004$) was the first malware designed for Symbian OS, and it used Bluetooth capabilities for infection \cite{ye2017survey}. Gemini ($2010$) was the first malware on the Android platform and was part of a mobile botnet system \cite{ye2017survey}. Since then, there has been an exponential growth of new malicious applications detected on Android OS (from $214,327$ in $2012$ to more than $4,000,000$ in $2018$) \cite{gdata}. Recently Symantec reported (February $2019$) every one in thirty-six mobile device has a high-risk application installed on it \cite{symantec}. According to McAfee threat report published in December $2018$, the current mobile malware growth is fueled by fake applications with an average infection rate of more than $10\%$ for every quarter in $2017$ \cite{mcafee}. Although there has been an aggressive growth of Android malware in the last decade, the dataset (malicious samples) available to the academic research community is limited \cite{ye2017survey} \cite{tam2017evolution}.

The primary defence against any malware attack is provided by Antivirus (AV) companies (Symantec, McAfee, Quick Heal, Kaspersky, etc.) and the anti-malware research community \cite{hicks2016exploratory} \cite{ye2017survey} \cite{sahay2020evolution} \cite{tam2017evolution}. Initially, AV engines were based on signature-based detection mechanism. A signature is a specific pattern (like known malicious instruction/activity sequence) maintained in an AV database \cite{sahay2020evolution}. However, the major drawbacks of the signature-based mechanism are that the approach is not scalable and is also suspectable to zero-day attacks \cite{ye2017survey} \cite{tam2017evolution}. Heuristic engines in AV often complement signature-based detection where malware experts define rules to detect malicious activities in an environment. Writing precise rules for such engines to identify malicious activity without increasing the false positive rate is a hard task \cite{ye2017survey}.

Recently researchers have begun exploring ways to develop a cutting edge malware detection system based on machine learning and deep learning \cite{hicks2016exploratory} \cite{ye2017survey} \cite{sewak2020deepintent} \cite{faruki2014android}. The above process requires data collection, feature extraction, feature engineering and building classification model. For example, Arp et al. constructed the Drebin dataset containing malicious Android applications \cite{arp2014drebin}. They extracted $5,45,000$ features with static analysis and built a support vector machine based malware detection model with an accuracy of $93.90\%$ \cite{arp2014drebin}. Static analysis conducts broad examination and if not performed carefully leads to a large feature vector, which then contributes to the curse of dimensionality and higher false-positive rate \cite{ye2017survey}. On the other hand, TaintDroid tracked information flow in Android APKs using dynamic analysis of thirty popular applications downloaded from third-party app stores \cite{enck2014taintdroid}. They found sixty-eight instances of potential misuse of the information in twenty different mobile apps. Dynamic analysis is often hard and infeasible to perform because of inadequate knowledge to trigger the malicious code path. Thus we used Android permissions gathered by static analysis to build the state-of-art malware detection system using machine learning and deep learning. The extensive list of permissions was handcrafted from the original Android documentation. Firstly we performed exploratory data analysis with correlation grid to gather useful insight about the data. Further, we performed a comprehensive feature vector analysis to reduce attributes using attribute subset selection methods (variance threshold) and also attribute creation methods (principal component analysis and auto-encoders). Finally, we used an extensive list of classifiers derived from the traditional set (decision tree, support vector machine, and k-nearest neighbour), ensemble methods (random forest and adaptive boosting) and three deep neural networks (shallow, deep and deeper) to build and compare different classification models. We performed an in-depth analysis to develop an efficient Android malware detection system and made the following contributions:

\begin{enumerate}
    \item We propose an Android malware detection system based on comprehensive feature engineering (using different attribute reduction techniques) followed by classification models (based on machine learning and deep neural network).
    \item Our baseline model built with the random forest classifier was able to achieve the highest accuracy ($94\%$). Our reduced feature models constructed with only $8\%$ Android permissions attained comparable accuracy with appreciable time-saving. To evaluate the efficacy of our approach, we used only $16$ Android permissions and achieved an accuracy of $93.3\%$ with random forest classifier ($\sim$ $1\%$ less than the baseline model). The decrease in Android permissions reduced the train and test time by half and tenth, respectively. This pattern of reduction in train and test time was virtually observed in almost all the analyzed classification models.
    \item Deep neural network models achieve comparable accuracy against machine learning models but have a massive computational penalty (hundred and ten times more train and test time respectively compared to random forest models). We also found that malicious applications tend to use similar Android permission sets while for benign applications, the set is more scattered.
\end{enumerate}

The rest of the paper is organized as follows. Section-2 presents a literature review of existing work on malware analysis and detection. Section-3 explains the background, broad framework and performance metrics. Section-4 discusses experimental setting and analysis (dataset, feature extraction, feature engineering and classification techniques) followed by experimental results and discussion. The last section concludes the paper by highlighting important points.
\section{Related Work}

Android malware detection is a rat-race\footnote{\url{https://attack.mitre.org/}} between malware developer and the anti-malware community. Currently, popular detection mechanisms like signature-based and heuristic-based detection have severe limitations \cite{hicks2016exploratory} \cite{ye2017survey} \cite{tam2017evolution}. Thus researchers are trying to develop state-of-the-art malware detection systems using machine learning and deep learning techniques. Building these systems is a two-stage process: attribute extraction and classification/clustering \cite{ye2017survey}. Android applications can be analyzed using the static or dynamic analysis to generate features for the construction of machine learning models. In static analysis, attributes are generated without running the code. However, in dynamic analysis, the sample application is executed in a sandbox to extract its behaviour. Then the extracted features are used to build models using classification/clustering algorithms for detecting malicious applications.

We have divided the literature review into two sub-areas:

(1) Feature extraction: An Android malware detection system named DroidDelver was developed by Hou et al. using API call blocks as the feature set on comodo cloud dataset \cite{hou2016droiddelver}. Lindorfer in Andrubis collected one million android applications from various sources and found that malicious applications tend to request more permissions ($12.99$) than benign applications ($4.5$) \cite{lindorfer2014andrubis}. In DroidMat, features like application components, intents, permissions, and API calls were extracted for building the detection model \cite{wu2012droidmat}. Drebin collected intents, application components, API calls, permissions and network address for malware detection \cite{arp2014drebin}. They concluded that certain combinations of hardware are requested more by malicious applications than benign ones. Sharma et al. grouped apps based on dangerous permissions and then used opcode to generate the feature vector \cite{sharma2018investigation}. Harris et al. developed a model for prediction of Android applications requesting excessive permissions. They also suggested changes in the way Android permissions are displayed and explained to the consumers \cite{harris2015mobile}. Nguyen et al. modified the Snapchat application by adding excessive permissions and repacking it. The modified application is then checked using static and dynamic analysis tools simulating a zero-day repacked malware attack \cite{nguyen2017exploitation}. Wang et al. extracted features viz. API calls and permissions from Android apps for DroidDeepLearner \cite{wang2016droiddeeplearner}. Zhou et al. discovered that $93\%$ of malware applications connect to the malicious source on the internet \cite{wang2016droiddeeplearner}. Sarma et al. found that malware apps tend to request network access more often than their benign counterparts \cite{sarma2012android}.

(2) Classification model: After constructing the feature vector, various algorithms can be used to build the malware detection system. Arp et al. with benchmark Drebin dataset used support vector machine on the feature set with $5,45,000$ attributes and achieved $93.8\%$ accuracy \cite{arp2014drebin}. However, the model suffers from the curse of dimensionality due to an enormous vector size. Further, Li et al. in $2018$ used the Drebin dataset to perform multilevel data pruning to find the $22$ most significant permissions and achieved an accuracy of $91.97\%$ \cite{li2018significant}. However, the cost of data pruning was not discussed in the paper, which might adversely affect the overall performance of the detection model. In the past, new permission(s) have been added in almost every major Android update. Thus, finding the new significant permissions for effective malware detection is a recursive exercise and should be cost-efficient. Also, MalPat ($2018$) combine permissions and API calls on the same dataset to obtain an F1 score of $98.24\%$ \cite{tao2017malpat}. MalPat used the top $50$ most sensitive APIs for classification, but again the cost of selection was not discussed in the article. Wang et al. constructed a deep belief network with one hidden layer called DroidDeepLearner and reported the highest accuracy of $92.67\%$ \cite{wang2016droiddeeplearner}. Sharma et al. used the tree-based classifiers viz. J48, random forest, functional trees, NBTree, \& logistic model tree and achieved the accuracy of $79.27\%$ with functional trees on Drebin dataset \cite{sharma2018investigation}. Later in $2016$, Hou et al. using the same dataset proposed DroidDelver. They analysed models based on support vector machine, deep learning, decision tree, naive bayes and obtained the highest accuracy of $94.04\%$ with decision tree model \cite{hou2016droiddelver}. Sewak et al. used the random forest classifier and deep neural network of various depths on Malicia project with different feature reduction techniques. They attained the highest detection accuracy of $99.78\%$ with random forest classifier \cite{sewak2018comparison}. DroidMat used k-means, naive bayes, k-nearest neighbour algorithms and accomplished the highest f-measure of $91.8\%$ \cite{wu2012droidmat}. Patri et al. performed entropy analysis and shapelet-based classification for PE files \cite{patri2017discovering}. Wenjia Li et al. in 2018 used kirin rules on Drebin to achieve a recall of $94.29\%$ \cite{li2018android}.
\section{Overview and Framework}

Machine learning and deep learning are subset of artificial intelligence and are currently used in various applications like email filtering, financial market analysis, information retrieval, computer vision and others. In the last decade, deep learning combined with current computational resources (hardware and software) has provided promising results in various fields like natural language processing, recommendation systems, medical image analysis and others. These models are often developed as a black box, thus have less interpretability. Explainability of any machine learning/deep learning model is a crucial factor for its real-world deployment today \cite{ye2017survey} \cite{rhue2019beauty}. In this paper, we performed an empirical investigation to build an Android malware detection system using machine learning and deep learning.

Presently Android OS enjoys the smartphone duopoly with a current market share of than $80\%$.  Principal protection for any user on the Android platform is derived from the Android permission system \cite{daniel2014strategies}. An Android application must request for the specific permission needed in order to access user data (SMS, photographs, etc.) or system resources like (storage, wifi, etc.). An application might be allowed to access the data or resource depending on the request. Based on the protection level, Google has classified Android permissions\footnote{\url{https://developer.android.com/guide/topics/permissions/overview}} into four different sets normal, dangerous, signature, and signatureOrSystem. We have used Android permissions to build a malware detection system using machine learning and deep neural network. Based on an exhaustive literature review and study of the Android platform, we were motivated to solve the following research questions:

\begin{enumerate}
    \item[$\bullet$] \textbf{Research Question-1}: Do malware designers use specific Android permission set to perform malicious activities? 
    \item[$\bullet$] \textbf{Research Question-2}: Can a reduced Android permission set be constructed for efficient Android malware detection?
    \item[$\bullet$] \textbf{Research Question-3}: Do Android malware detection model(s) based on machine learning tend to use less computational resources as compared to the deep learning algorithms?
\end{enumerate}

\subsection{Framework Design}

\begin{figure}[!ht]
	\centering
	\includegraphics[width=1.0\linewidth]{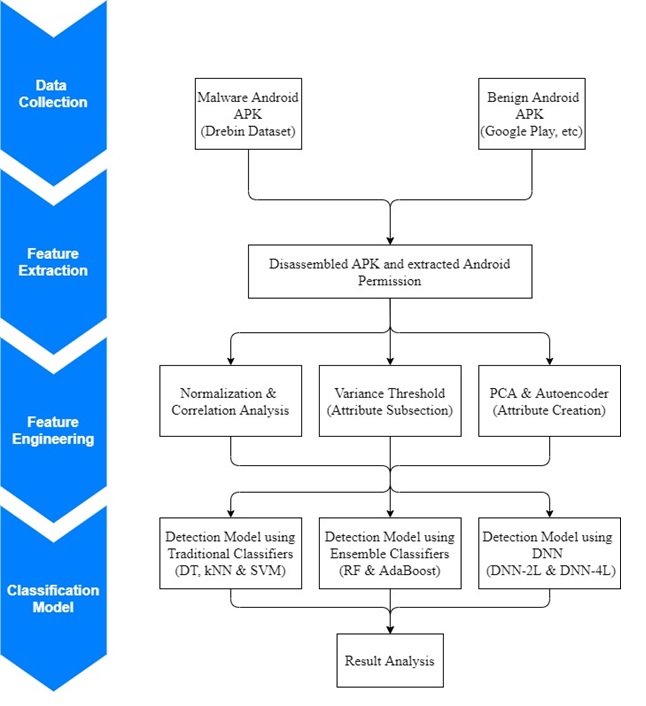}
	\centering
	\caption{Procedure for the construction of an efficient Android malware detection system}
	\label{framework}
\end{figure}

This study aims to propose an effective and efficient Android malware detection system. Figure \ref{framework} shows the systematic framework used for empirical analysis for malware classification. The design is divided into four sub-modules:

\begin{enumerate}
    \item \textbf{Data collection}: We gathered more than $10,000$ Android applications (benign and malware) from various sources for our experimental analysis.
    \item \textbf{Feature extraction}: Features are the backbone of any machine learning solution. We extracted Android permissions from the applications to be used as features for our models.
    \item \textbf{Feature Engineering}: It is the process of understanding features using domain knowledge and statistical tools to improve the effectiveness and performance of a classification model. We used correlation analysis followed by feature reduction methods like variance threshold, principal component analysis and auto-encoders to reduce the size of the Android permission set used for building the malware detection models to avoid the curse of dimensionality.
    \item \textbf{Classification Models}: Various machine learning and deep learning algorithms can be used to build the detection models. We used traditional classifiers, ensemble methods and deep neural network algorithms to build an efficient Android malware detection system. All the above steps are discussed in-depth later in the paper.
\end{enumerate}

\subsection{Performance Metrics}

A classification model is evaluated on the following four values (\textit{True Positive} (TP), \textit{False Positive} (FP), \textit{True Negative} (TN), and \textit{False Negative} (FN). From these predicted values, various metrics can be derived to measure and understand the performance of the classification model.

\begin{enumerate}

\item Accuracy: It is the ratio of the correct predictions to the total number of predictions.
\begin{eqnarray}
	Accuracy &=& \frac{TP + TN}{TP + FP + TN + FN}
\end{eqnarray}

\item Recall (TPR): The Recall is the ratio of correctly predicted positive values to the total actual positive values.
\begin{eqnarray}
	Recall &=& \frac{TP}{TP + FN}
\end{eqnarray}

\item Specificity (TNR): TNR is the ratio of the number of correctly classified benign apps to the total number of benign apps.
\begin{eqnarray}
	TNR &=& \frac{TN}{TN + FP}
\end{eqnarray}

\item Receiver Operating Characteristic (ROC): It illustrates the separability of classes in a model, which is the ability of a model to classify malicious application as malware and benign application as benign. The ideal model has AUC value at $1$.

\end{enumerate}

\section{Experimental Setting and Analysis}

This section discusses the dataset (malware and benign), the process of feature extraction, feature engineering methods performed on the dataset and different classification algorithms used to build the efficacious Android malware detection system.

\subsection{Malware and Benign Datasets}

Input data quality holds the vital key for building any effective classification model. Daniel Arp and others compiled the Drebin dataset containing $5,560$ Android malware samples downloaded from the Google Play store and various other sources \cite{arp2014drebin}. It also contains all the malicious applications from the Android Malware Genome Project \cite{zhou2012dissecting}. Drebin includes apps from multiple malware families like \textit{FakeInstaller}, \textit{DroidKungFu}, \textit{GoldDream}, and \textit{GingerMaster}. After reviewing the different model proposed by various authors, we have also used the benchmark Drebin dataset as the representative of malicious Android applications for all our experiments \cite{arp2014drebin} \cite{ye2017survey} \cite{li2018significant}.

Google Play Store\footnote{\url{https://play.google.com/store}} and other third-party apps stores are the primary distributors of Android applications (.apk) to users. For obtaining the benign samples, we downloaded $\sim8,000$ Android apps from the Google Play store. To validate whether the downloaded apps are malicious or not, we used the services of VirusTotal\footnote{\url{https://www.virustotal.com/gui/home/upload}} (a subsidiary of Alphabet Inc., which aggregates result of many AV products and online search engines). The reports generated by VirusTotal were used to segregate the benign samples for our experiments. An Android application is labelled as benign only if all the AVs from VirusTotal declare it as non-malicious. The remaining samples were non benign and hence discarded. Thus the final dataset for our analyses contained $5,560$ malware and $5,721$ benign samples.

\subsection{Feature Extraction}

Features can be extracted by static/dynamic analysis of Android applications to build a classification model. Essentially the feature vector is a fundamental pillar of any malware detection system. In this project, Android applications were disassembled using a reverse engineering software called Apktool\footnote{\url{https://ibotpeaches.github.io/Apktool/}}. A disassembled app contains AndroidManifest.xml, smali files, library files, assets, etc. The list of Android permissions to be used by an application are declared in the AndroidManifest.xml file. We handcrafted the extensive list of all the permissions using original Android documentation \footnote{\url{https://developer.android.com/reference/android/Manifest.permission}}. The list contains $197$ permissions and their respective API level. We have considered all the permissions from API level $1$ to API level Q($29$) (launched in 2019). The list also contains depreciated permissions like \textit{PERSISTENT\_ACTIVITY} in API level $15$, \textit{GET\_TASKS} in level $21$, etc. which cannot be used in the newer version of Android. Finally, all the apps from malware and the benign dataset were decompiled using Apktool. Few Android applications were corrupted and thus were discarded.  Lastly, the parser scanned through AndroidManifest.xml of each application to generate the feature vector containing the list of permissions used by that application. Thus the final feature vector representing the dataset is $11,274 \times 197$, where a row represents a particular Android app (malware or benign), and a column represents Android permission.

\subsection{Feature Engineering}

\begin{figure}[!ht]
	\centering
	\includegraphics[width=1.0\linewidth]{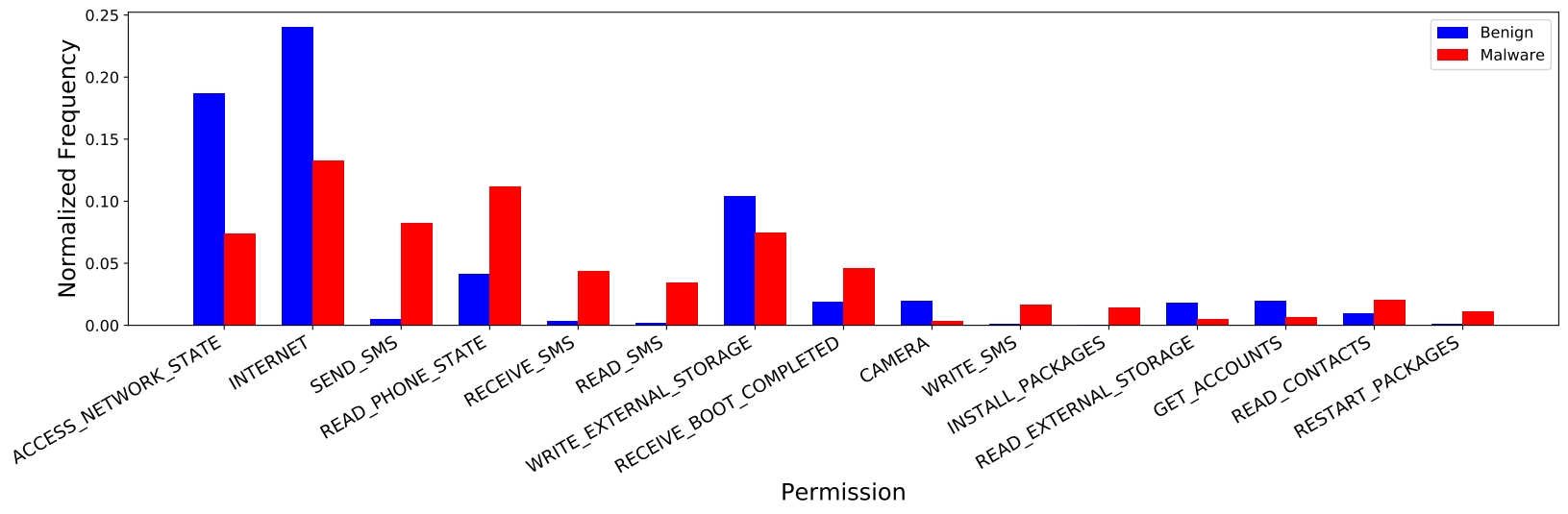}
	\centering
	\caption{Normalized frequency of the permissions in malicious and benign applications}
	\label{NF}
\end{figure}

After rigorous feature extraction from Android applications, the next step was to perform feature engineering and to develop the classification models. Initial analysis of the feature vector revealed some useful insights into the usage of the Android permissions. In total, $59$ permissions have never been used by any applications (malware/benign) in the dataset. Individually the number is $82$ and $67$ for malware and benign applications respectively. Analyzing further, some applications were using deprecated permissions (like \textit{RESTART\_PACKAGES} in API level $15$, \textit{GET\_TASKS} in level $21$,  etc.) which cannot be used in the newer versions of Android. Also, the distribution of permission usage is not uniform across all applications. Some permissions are heavily used by malicious applications while others by benign applications. Figure \ref{NF} shows the detailed visualization of permissions and their normalized frequency in the dataset. Android permissions like \textit{ACCESS\_NETWORK\_STATE}, \textit{INTERNET} and \\\textit{WRITE\_EXTERNAL\_STORAGE} are used by both benign and malware samples but extensively by benign applications. Also, permissions like \textit{SEND\_SMS}, \textit{RECEIVE\_SMS}, \textit{READ\_SMS}, \textit{WRITE\_SMS} and \textit{INSTALL\_PACKAGES} are heavily used by malicious applications but rarely by benign applications.

\begin{figure}[!ht]
	\centering
	\includegraphics[width=0.8\linewidth]{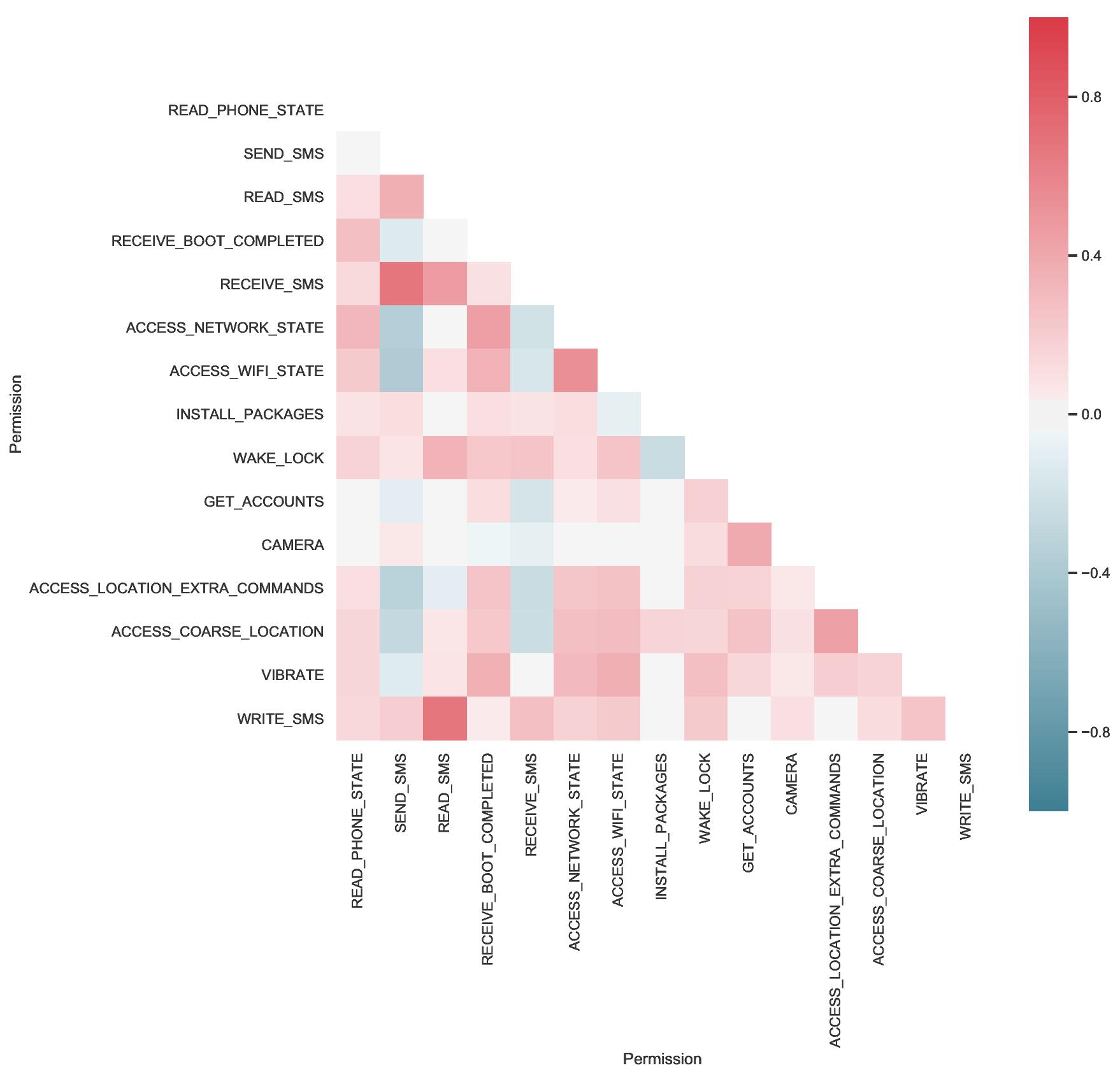}
	\centering
	\caption{Correlation matrix of permissions used in the malicious applications}
	\label{corr}
\end{figure}

Correlation grid describes the degree of linear relationship between two variables. Figure \ref{corr} shows the correlation between the top fifteen features (Android permissions) derived from the random forest in malicious applications. \textit{READ\_SMS} \& \textit{WRITE\_SMS} and \textit{SEND\_SMS} \& \textit{RECEIVE\_SMS} has the highest correlations of $0.68$ followed by \textit{ACCESS\_NETWORK\_STATE} \& \textit{ACCESS\\\_WIFI\_STATE}, \textit{READ\_SMS} \& \textit{RECEIVE\_SMS}, \textit{ACCESS\_NETWORK\_STATE} \& \textit{RECEIVE\_BOOT\_COMPLETED} of $0.53$, $0.47$ and $0.45$ respectively. \textit{READ\\\_PHONE\_STATE} is often used by malware to gather phone number, current cellular network information and personal information like call list, SMS etc. \textit{SEND\_SMS} is also used by malware to subscribe to the victim to unwanted paid services. All these permissions are used in tandem with each other to perform the desired malicious task.

\subsection{Feature Reduction}

As the number of unique system-defined Android permissions are $197$, thus the feature vector also consists of $197$ attributes. A classification model that builds on a large number of attributes will be computationally expensive, have higher train/test time, be less interpretable and is more likely to suffer from the curse of dimensionality. Thus, we performed feature reduction using both feature subset selection method (variance threshold) and attribute creation techniques (principal component analysis and autoencoder).

\subsubsection{Variance Threshold (VT):}

\begin{figure}[!ht]
	\centering
	\includegraphics[width=.70\linewidth]{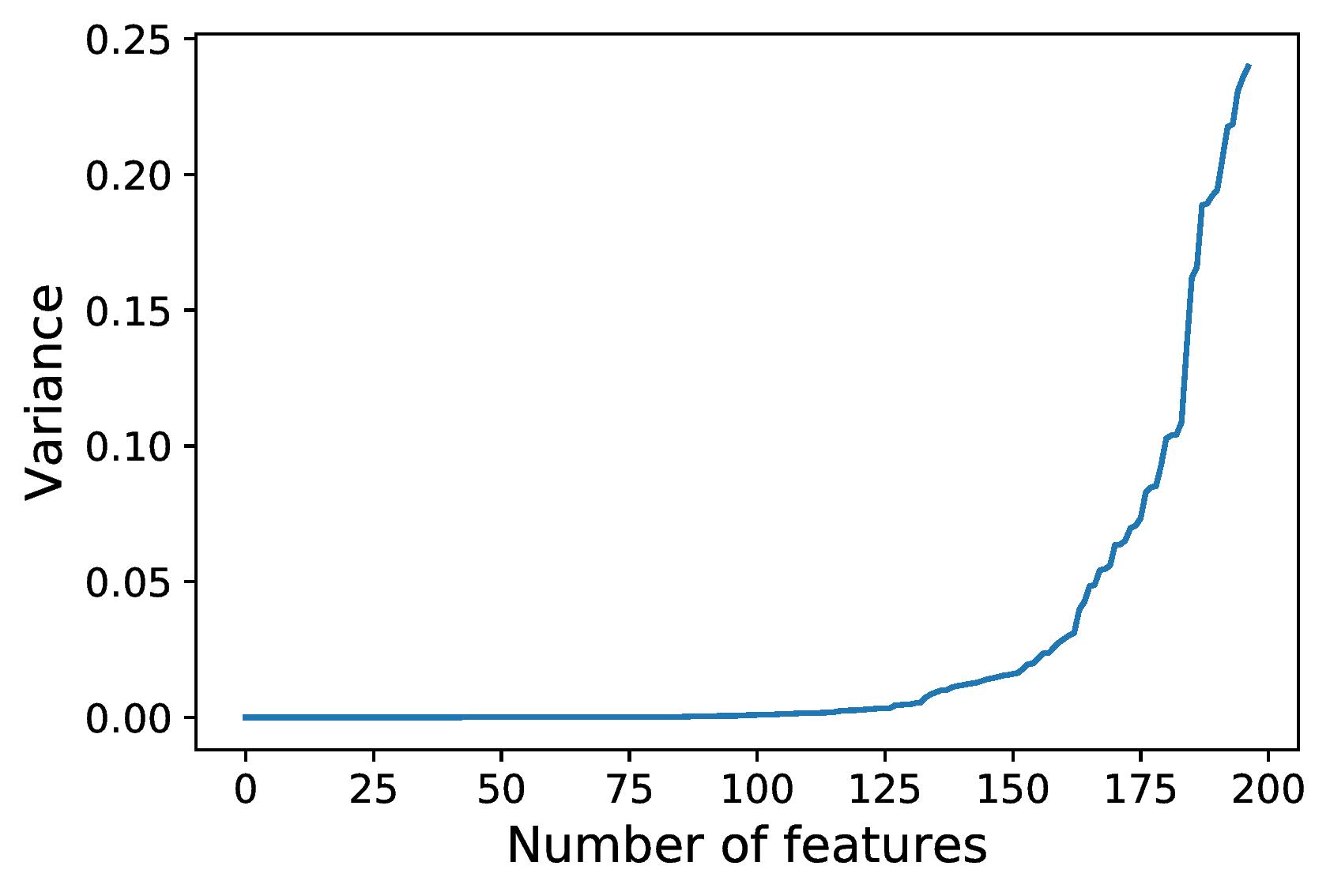}
	\centering
	\caption{Variance in Android permissions}
	\label{vt}
\end{figure}

VT is a feature reduction technique in which the variance of each feature in a dataset is calculated and features having less variance (below a threshold value) are dropped from the final vector. Figure \ref{vt} shows the plot between the features and corresponding variance values in the dataset. The plot clearly shows many permissions have very low variance, and thus have less predictive power during classification. Fifty-nine permissions have zero variance, and hence they do not contribute to the classification. Ten permissions have variance less than $0.0001$ and remaining features have variance between $0.0001$ and $0.24$. Permissions \textit{READ\_PHONE\_STATE}, \textit{WRITE\_EXTERNAL\_STORAGE}, and \textit{ACCESS\_WIFI\_STATE} have the highest variance of $0.24$, $0.23$, and $0.23$ respectively. In Figure \ref{vt}, an elbow is visible at $0.10$ (set as threshold value), and thus the final VT feature vector consists of only $16$ permissions. Table \ref{permission_table} list all the $16$ Android permissions arranged in descending order based on the variance.

\begin{table}[!ht]
	\centering
	\caption{Significant Permission Ranking}
	\label{permission_table}
    \begin{tabular}{|c|c|}
    \hline
    Android Permission       & \textbf{Ranking} \\ \hline
    READ\_PHONE\_STATE       & 1                \\ \hline
    WRITE\_EXTERNAL\_STORAGE & 2                \\ \hline
    ACCESS\_WIFI\_STATE      & 3                \\ \hline
    RECEIVE\_BOOT\_COMPLETED & 4                \\ \hline
    WAKE\_LOCK               & 5                \\ \hline
    SEND\_SMS                & 6                \\ \hline
    ACCESS\_COARSE\_LOCATION & 7                \\ \hline
    ACCESS\_NETWORK\_STATE   & 8                \\ \hline
    ACCESS\_FINE\_LOCATION   & 9                \\ \hline
    VIBRATE                  & 10               \\ \hline
    RECEIVE\_SMS             & 11               \\ \hline
    READ\_SMS                & 12               \\ \hline
    READ\_CONTACTS           & 13               \\ \hline
    GET\_ACCOUNTS            & 14               \\ \hline
    WRITE\_SMS               & 15               \\ \hline
    CHANGE\_WIFI\_STATE      & 16               \\ \hline
    \end{tabular}
\end{table}

\subsubsection{Principal Component Analysis (PCA):}

\begin{figure}[!hb]
	\centering
	\includegraphics[width=0.8\linewidth]{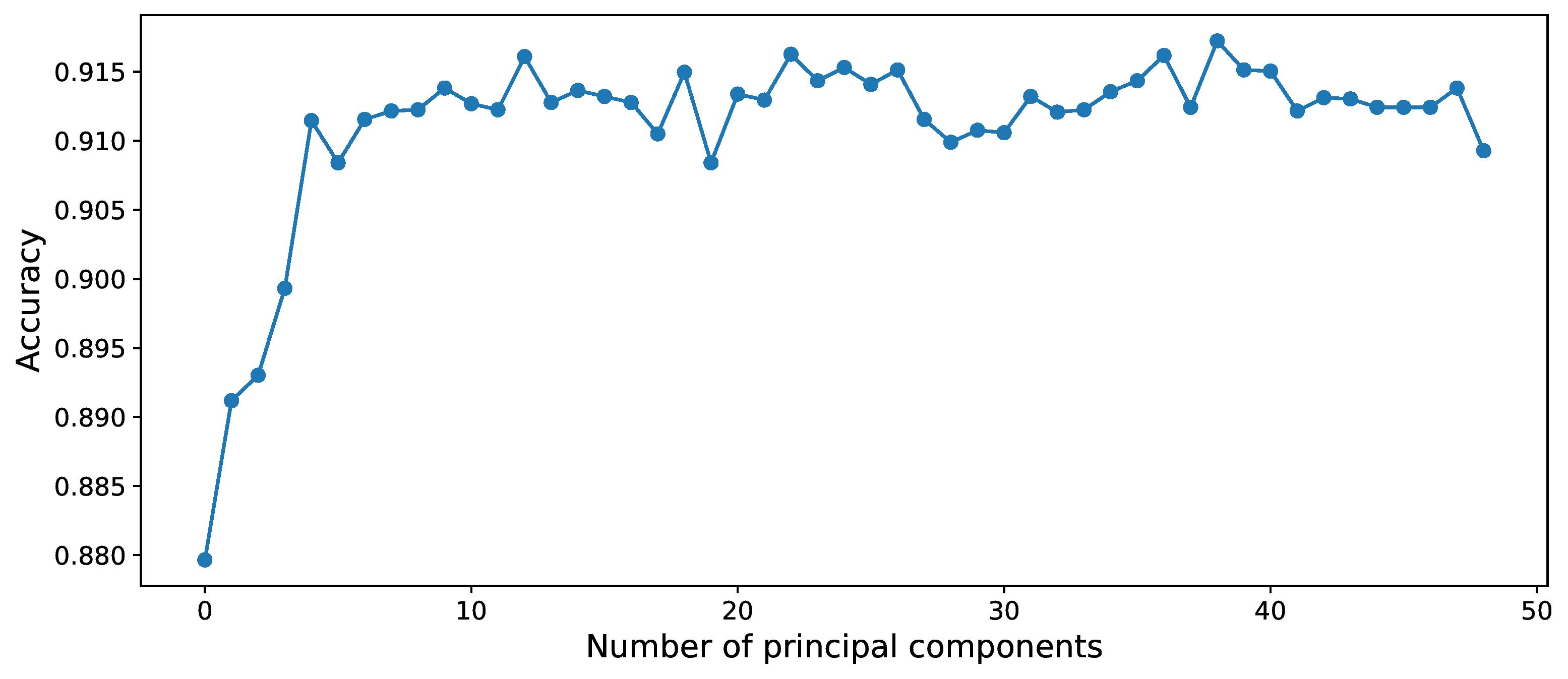}
	\centering
	\caption{Principal component analysis for Android permissions}
	\label{pca}
\end{figure}

PCA is another feature reduction technique where the dimensionality of a dataset is reduced by transforming the data to new orthogonal coordinate axes such that loss of information and variance is minimum. Further, highly correlated dimensions are dropped at the cost of a small decrease in the accuracy to reduce the curse of dimensionality. Figure \ref{pca} shows the model accuracy with the number of principal components by random forest classifier. Initially, as the principal components are added there is a sudden increase in the model accuracy, which stabilizes around $16$ principal components. Hence top $16$ principal components are chosen to represent the data with PCA feature reduction technique. Since PCA is an attribute creation method, thus a  particular principal component can represent one or more Android permissions.

\subsubsection{Auto Encoder (AE):}

AE uses deep learning to reduce the dimensionality of a dataset. AE takes the set of features as the input and produces a reduced set as the output. AE consists of an encoder (input layer to code layer) which maps the input data to reduced latent space, and a decoder (code layer to the last layer) which tries to reconstruct the input data back. The number of neurons in the hidden layer(s) in AE is decreased gradually during encoding and increased again during decoding. AE is first trained on a complete network after which decoder is discarded, and the only encoder is used for feature reduction. We have designed a shallow and deep AE to achieve two distinct data transformations on the dataset.

\begin{itemize}
    \item[$\bullet$] \textbf{AE with 1 Layer (AE-1L)} is a shallow autoencoder with only one hidden layer/code layer with $64$ neurons. The design of AE-1L is $197$-$64$-$197$.
    \item[$\bullet$] \textbf{AE with 3 Layer (AE-3L)} is a deep autoencoder with three encoding layers having $64$, $32$ and $16$ neurons consecutively. The overall design of AE-3L is $197$-$64$-$16$-$64$-$197$.
\end{itemize}

\begin{figure}[htbp]
	\centering
	\begin{minipage}[b]{0.49\textwidth}
		\includegraphics[width=1.0\linewidth]{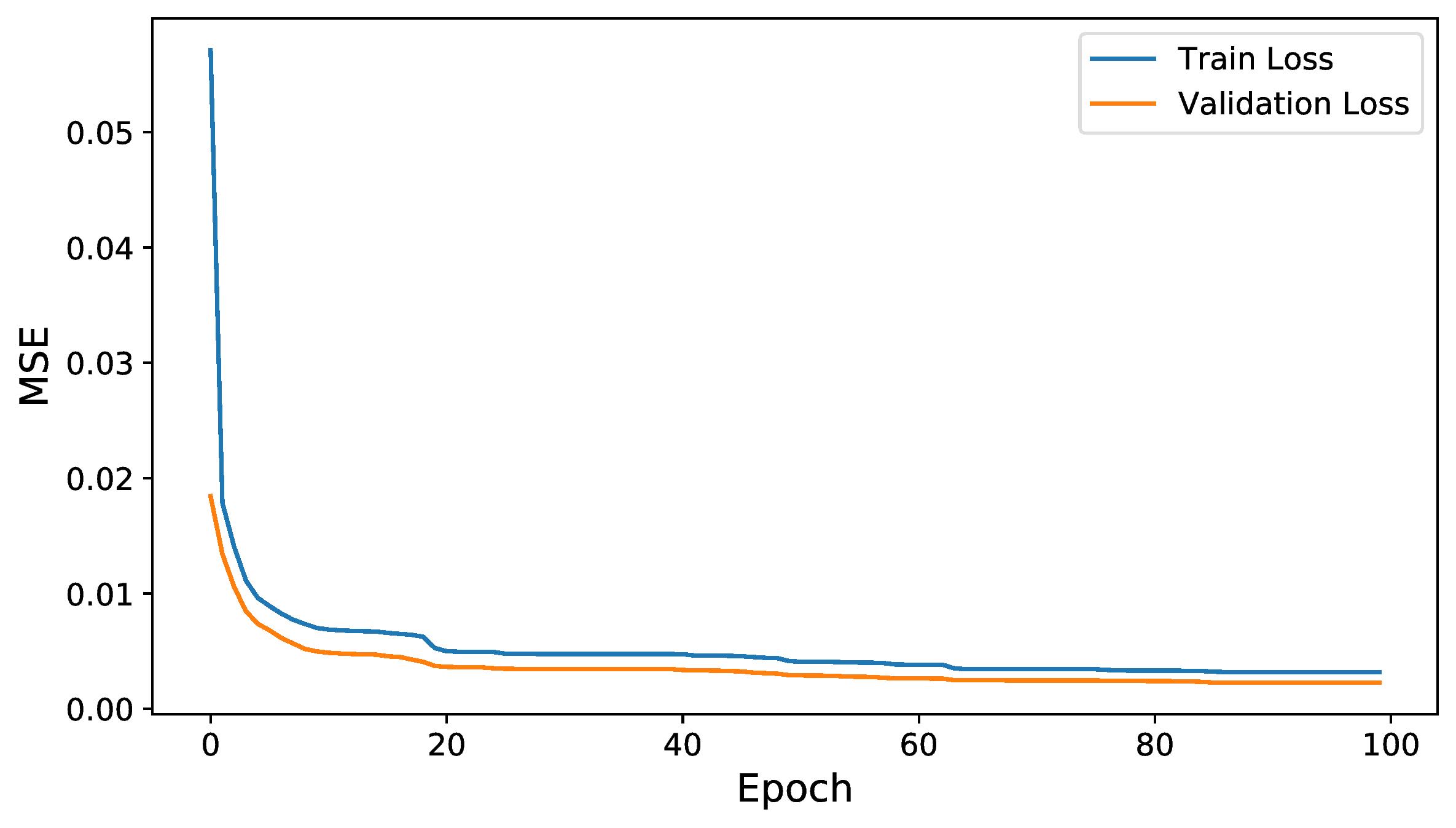}
		\caption{Training and validation loss at different epochs for AE-1L}
		\label{AE1LMSE}
	\end{minipage}
	\hfill
	\begin{minipage}[b]{0.49\textwidth}
		\includegraphics[width=1.0\linewidth]{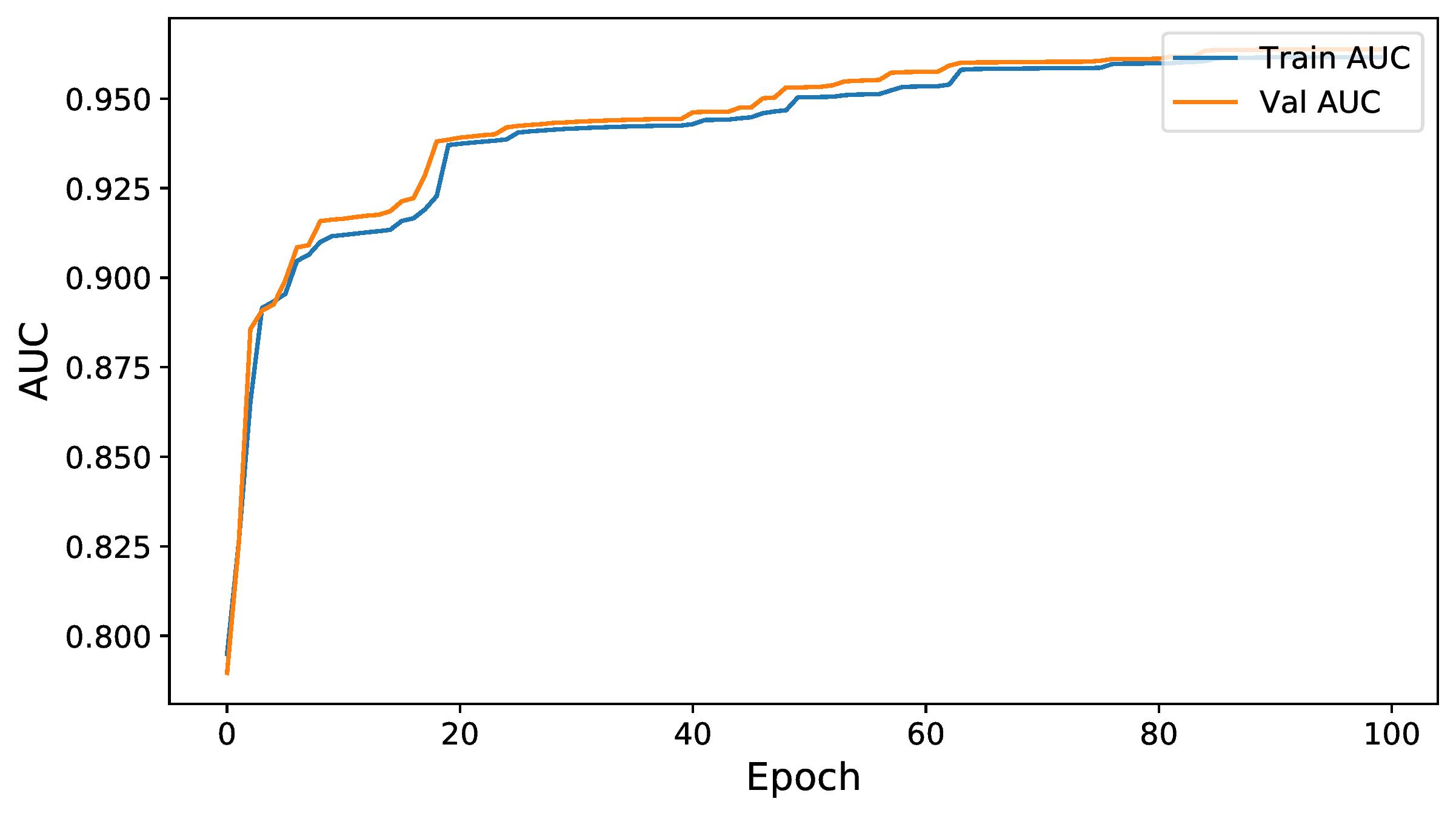}
		\caption{Training and validation AUC at different epochs for AE-1L}
		\label{AE1LAUC}
	\end{minipage}
\end{figure}

By design, all the layers (encoder and decoder) in both the AEs are fully connected with Rectified Linear Unit (ReLU) activation function except the code layer, which uses the sigmoid function because of the binary data. A dropout of $0.4$ is set at all the layers for better generalization and to avoid overfitting. Adam optimizer is used to train both the AE with mean square error (MSE) as the loss function. The training and validation split of $80:20$ is maintained, and both the AE are trained for $80$ epochs with a batch size of $64$. After training, Keras is used in the backend to derive an output from the code layer. Figure \ref{AE1LMSE} and figure \ref{AE1LAUC} shows the mean square error and AUC for the training and validation data over different epochs for AE-1L. During initial cycles, both MSE and AUC are jittery but become stable after training the AE-1L for $80$ epochs. It also signifies that both the AEs are trained well and are not over-fitting or under-fitting the dataset. 

Finally, Table \ref{small_table} shows the different feature reduction techniques and the number of output features produced by each method. Original data contains all the $197$ features without any feature reduction. VT performs attribute sub-selection, and top $16$ Android permissions as features were selected based on variance for the final vector. Attribute creation method like PCA, AE-1L and AE-3L have created the new feature set of size $16$, $64$ and $16$ respectively.

\begin{table}[!ht]
	\centering
	\caption{Feature reduction technique and corresponding reduced vector size}
	\label{small_table}
	\begin{tabular}{|c|c|}
		\hline
		\textbf{Feature Reduction} 	& \textbf{Number of features} \\ \hline
		OD\rlap{\textsuperscript{1}}			              	& 197                     \\ \hline
		VT\rlap{\textsuperscript{2}}                         	& 16                      \\ \hline
		PCA\rlap{\textsuperscript{3}}                     	& 16                      \\ \hline
		AE-1L\rlap{\textsuperscript{4}}             			& 64 (code layer)   \\ \hline
		AE-3L\rlap{\textsuperscript{5}}               		& 16 (code layer)   \\ \hline
	\end{tabular}
	{\\\scriptsize\textsuperscript{1}Original Data}
	{\scriptsize\textsuperscript{2}Variance Threshold}
	{\scriptsize\textsuperscript{3}Principal Component Analysis\\}
	{\scriptsize\textsuperscript{4}AE with 1 Layer}
	{\scriptsize\textsuperscript{5}AE with 3 Layers}
\end{table}

\section{Malware Detection Models}

We used three different categories of classifiers for building a permission-based Android malware detection system. The first group contain traditional classifiers like Decision Tree (DT), Support Vector Machine (SVM), and k-Nearest Neighbour (kNN). The second group include ensemble methods like Random Forest (RF) and Adaptive Boosting (AdaBoost). The third group consists of Deep Neural Network (DNN) of different depths.

\subsection{Detection Model using Traditional Classifiers}
DT is a tree-based supervised learning algorithm which maximizes the information gain at every decision node. Gini impurity was used as the support criteria to choose the best split at each stage. There was no limit on the maximum depth of the tree, but the minimum number of samples at each leaf node was set to five to avoid overfitting. kNN classifies data based on its nearest k-neighbours in feature vector space. We considered five nearest neighbours to classify the data point with each neighbourhood weighted equally. SVM uses hyperplane and support vector to classify data into different classes. It is also called the maximum margin classifier since it tries to maximize the margin between the data and the hyperplane. We used LinearSVC kernel for classification, and decision function shape was one vs rest. The penalty parameter was set to \textit{l1} since the feature vector space was sparse.

\subsection{Detection Model using Ensemble Classifiers}
RF is also a supervised machine learning algorithm, which is an ensemble of a large number of DTs with different parameters. The class which is predicted by the majority of trees is the final output. Due to a large number of independent trees, the error of an individual tree does not affect the overall outcome. RF also perform feature sub-selection to avoid overfitting. The number of DTs in our model was $100$ with Gini impurity as the split criteria. Bootstrapping and Out-of-bag was set to true for better generalization. There was no limit on the maximum depth of the tree, but the minimum number of samples to the leaf node was set to five. AdaBoost uses a series of weak classifiers for the final classification. The base estimator (weak classifiers) was DT with a maximum depth of one. \textit{SAMME.R} was used to boost the results of the base estimators. The number of estimators was set to $50$, and the learning rate was $1$.

\subsection{Detection Model using DNN}

DNN is a supervised learning technique which uses deep learning to build classification models. We have built three different DNN models with one hidden layer (shallow network), three hidden layers and five hidden layers (deep network) with the following design:

\begin{itemize}
    \item[$\bullet$] \textbf{DNN with 2 Layers (DNN-2L)} has one input layer and a hidden layer containing $197$ and $64$ neurons, respectively. The output layer acts as the decision point and has only one neuron. The design of DNN-2L is $197$-$64$-$1$.
    \item[$\bullet$] \textbf{DNN with 4 Layers (DNN-4L)} has one input layer of $197$ neurons, followed by three hidden layers containing $128$, $32$, and $8$ neurons, respectively. The design of DNN-4L is $197$-$128$-$32$-$8$-$1$.
    \item[$\bullet$] \textbf{DNN with 7 Layers (DNN-7L)} is a deep network with six hidden layers containing $128$, $64$, $32$, $16$, $8$, and $4$ neurons, respectively. The design of DNN-7L is $197$-$128$-$64$-$32$-$16$-$8$-$4$-$1$.
\end{itemize}

\begin{figure}[htbp]
	\centering
	\begin{minipage}[b]{0.49\textwidth}
		\includegraphics[width=1.0\linewidth]{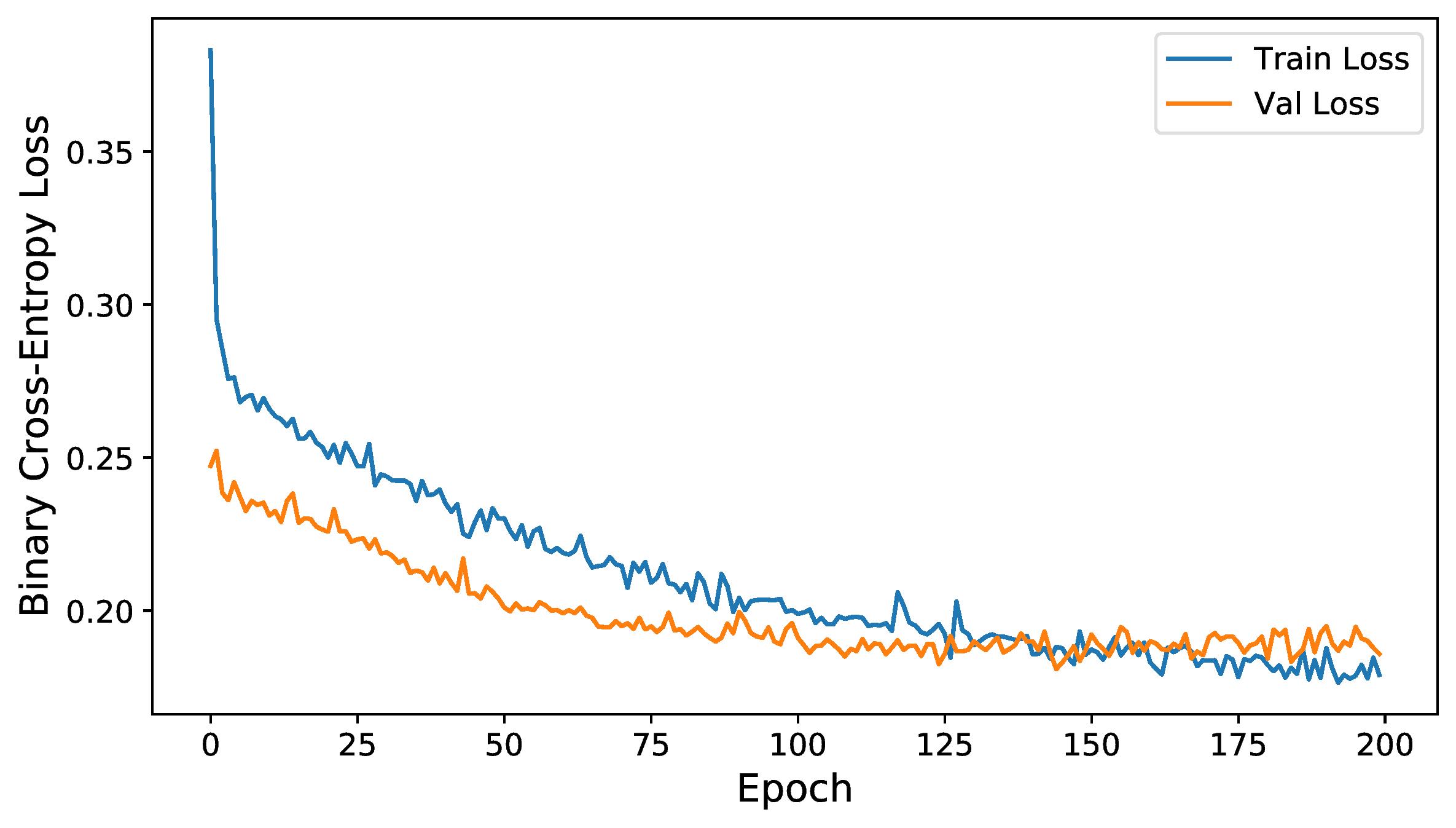}
		\caption{Training and validation loss at different epochs for DNN-4L}
		\label{NN4LBCE.jpg}
	\end{minipage}
	\hfill
	\begin{minipage}[b]{0.49\textwidth}
		\includegraphics[width=1.0\linewidth]{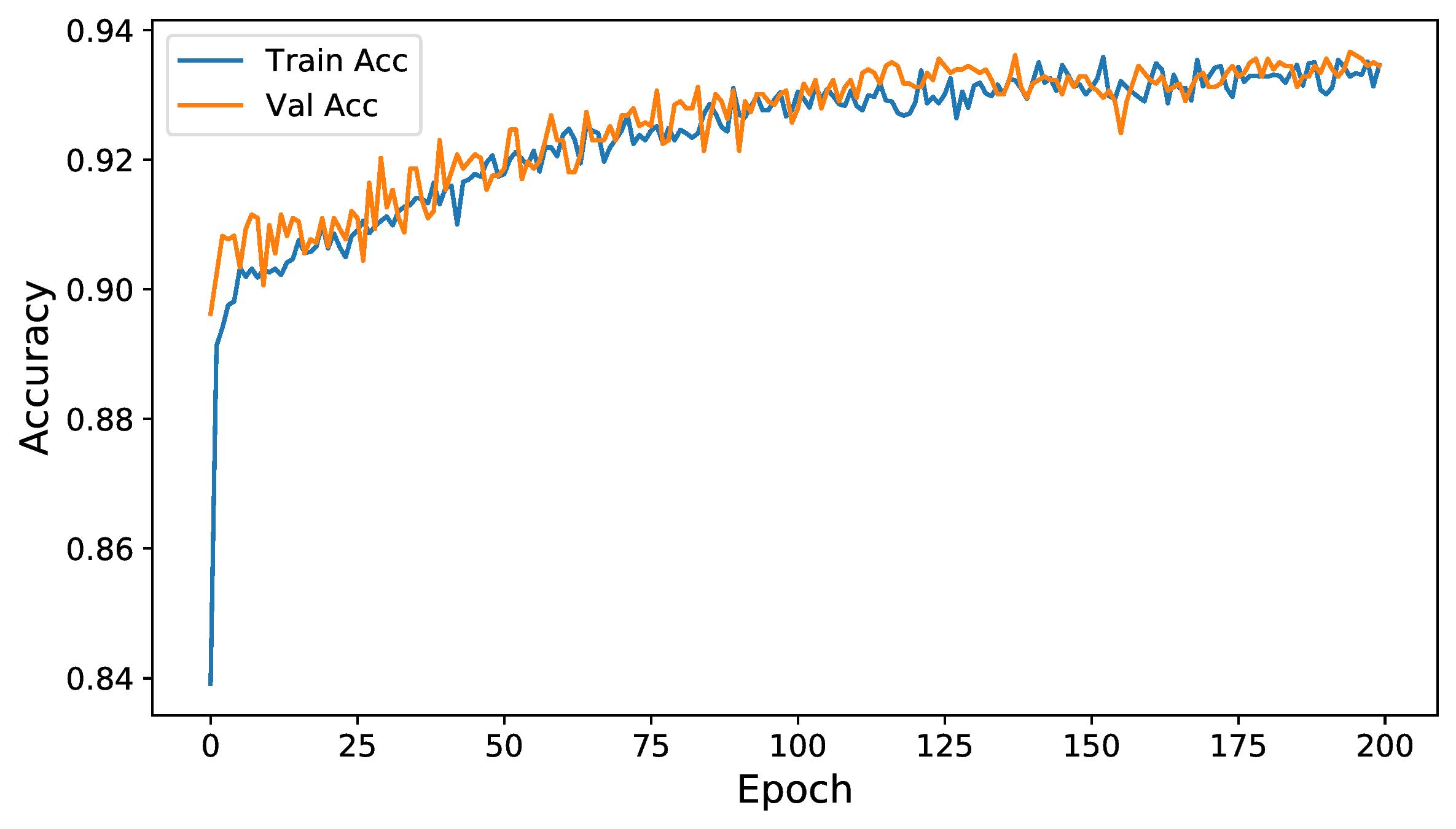}
		\caption{Training and validation AUC at different epochs for DNN-4L}
		\label{NN4LAcc}
	\end{minipage}
\end{figure}

All the DNNs are fully connected network with ReLU as activation function at all the layers except the output layer, which has a sigmoid function. A split ratio of $80:20$ for training and validation set is used to train the model. Again, the dropout rate of $0.4$ is used to avoid overfitting of DNN models. For training, Adam optimizer with a learning rate of $0.1$ and binary cross-entropy loss function is used. The DNNs are trained over $150$ epochs with a batch size of $64$. Figure \ref{NN4LBCE.jpg} and figure \ref{NN4LAcc} shows binary cross-entropy loss and AUC values for training and validation set of DNN-4L with VT data. The figures clearly show that the DNN-4L become stable after training for $150$ epochs and does not over-fit or under-fit the dataset.

\subsection{Results and Discussion}

Table \ref{big_table} shows the performance of selected classification models with different feature reduction techniques. Random forest models produced most balanced classifications results (highest AUC) with original data ($98.1$), VT data ($97.0$), AE-1L data ($97.7$), PCA data ($97.6$) and AE-3L data ($97.2$). On the other hand, classification models based in kNN tended to overfit the malware class.

In terms of Accuracy (Acc) RF with the original data achieved the highest score of $94.0$, followed by RF with VT data at $93.3$. In neural network models, both DNN-4L (VT data) and DNN-2L (AE-1L) achieved a high accuracy score of $93.1$. kNN (original data) and kNN (VT data) reported the lowest accuracy of $78.7$ and $78.4$ respectively since they were overfitting the malware class. As far as TPR is a concern, kNN (original data) and kNN (VT data) achieved the highest rates of $97.8$ and $96.4$, respectively. It also demonstrates that malicious applications use similar permission sets. The lowest TNR has also been provided by kNN with original and VT data at $59.7$ and $60.4$ respectively, which implies that benign apps use a diverse set of permissions. Highest TNR was recorded for DNN-7L (AE-1L data) and DNN-7L (original data) at $95.8$ and $95.4$ respectively, which is skewed towards the benign class.

In general, lower training time was reported for kNN, DT and RF with less number of features like VT data ($16$ features), PCA data ($16$ features), and AE-3L data ($16$ features). Also, there is a considerable reduction in train and test time when the model is trained on smaller feature vectors (VT, PCA and AE-3L) without much penalty on model accuracy. In the case of RF, train time was reduced to half for smaller vectors (VT data) with less than $1\%$ penalty on the model accuracy. It signifies that only $16$ features from VT, PCA and AE-3L are enough to represent that data and can be effectively used to build the classification model with less training and test time. Since kNN models store the feature vector with minimal preprocessing, thus they have considerably less training time. On the other hand, all DNNs have extremely large training time for building classification models. Although training time is a one-time activity in case of malware detection, the model needs to be updated quite frequently based on current telemetry. Similarly, reducing the feature vector has a positive impact on test time as well, where DT and RF outperformed other classifiers. Test time is particularly relevant because AV engines might have to run throughout $24 \times 7$ for detecting malicious activities.

Table \ref{compare_table} shows the detection rate of our proposed model with ten popular AVs. Our model w/RF achieved a higher detection rate of $93.3\%$ with only $16$ Android permissions which is much higher than most AV engines. On the other hand, Drebin used $5,45,000$ features to achieve $93.90\%$ detection rate with SVM, but their model suffered from the curse of dimensionality. Li et al. achieved detection rate of $91.36\%$ with $22$ features with a computationally expensive three-stage pruning approach \cite{li2018significant}.

\begin{table}[!htbp]
	\caption{Malware Detection Rates (Our model vs Antivirus)}
	\label{compare_table}
\begin{tabular}{|c|c|c|c|c|c|c|c|c|c|c|c|}
\hline
\begin{tabular}[c]{@{}c@{}}Our\\ w/RF\end{tabular} & \begin{tabular}[c]{@{}c@{}}Our\\ w/D4L\end{tabular} & AV1   & AV2   & AV3   & AV4   & AV5   & AV6   & AV7   & AV8  & AV9  & AV10 \\ \hline
93.3                                                     & 93.1                                                         & 88.22 & 83.62 & 83.60 & 83.28 & 81.57 & 56.02 & 19.28 & 7.93 & 7.91 & 4.22 \\ \hline
\end{tabular}
\end{table}

\begin{table}[htbp]
\caption{Performance of multiple classifiers with different feature reduction techniques}
\label{big_table}
\centering
\begin{tabular}{|c|c|c|c|c|c|c|}
\hline
\textbf{Feature Reduction} & \textbf{Classifier} & \textbf{Accuracy} & \textbf{TPR}  & \textbf{AUC}  & \textbf{\begin{tabular}[c]{@{}c@{}}Train Time\\ (sec)\end{tabular}} & \textbf{\begin{tabular}[c]{@{}c@{}}Test Time\\ (sec)\end{tabular}} \\ \hline
OD\rlap{\textsuperscript{1}}                         & DT\rlap{\textsuperscript{6}}                  & 92.6              & 91.9          & 94.5          & 0.052                     & 0.003                    \\
VT\rlap{\textsuperscript{2}}                         & DT                  & 92.3              & 90.9          & 94.7          & 0.012                     & \textbf{0.001}           \\
PCA\rlap{\textsuperscript{3}}                        & DT                  & 91.2              & 90.7          & 92.6          & 0.059                     & 0.002                    \\
AE-1L\rlap{\textsuperscript{4}}                      & DT                  & 91.4              & 91.1          & 92.7          & 0.258                     & 0.003                    \\
AE-3L\rlap{\textsuperscript{5}}                      & DT                  & 90.5              & 89.8          & 91.6          & 0.476                     & 0.002                    \\ \hline
OD                         & kNN\rlap{\textsuperscript{7}}                 & 78.7              & \textbf{97.8} & 84.2          & 0.133                     & 4.044                    \\
VT                         & kNN                 & 78.4              & 96.4          & 84.5          & 0.080                     & 0.421                    \\
PCA                        & kNN                 & 91.4              & 91.3          & 95.7          & \textbf{0.005}            & 0.119                    \\
AE-1L                      & kNN                 & 90.8              & 93.0          & 95.4          & 0.100                     & 1.950                    \\
AE-3L                      & kNN                 & 88.4              & 90.3          & 94.7          & 0.084                     & 0.327                    \\ \hline
OD                         & SVM\rlap{\textsuperscript{8}}                 & 91.0              & 89.4          & 96.3          & 5.634                     & 0.929                    \\
VT                         & SVM                 & 89.1              & 87.3          & 95.2          & 1.186                     & 0.116                    \\
PCA                        & SVM                 & 87.3              & 87.6          & 94.1          & 0.624                     & 0.088                    \\
AE-1L                      & SVM                 & 89.5              & 86.9          & 95.4          & 6.372                     & 1.061                    \\
AE-3L                      & SVM                 & 86.6              & 86.2          & 93.0          & 7.200                     & 1.185                    \\ \hline
OD                         & RF\rlap{\textsuperscript{9}}                  & \textbf{94.0}     & 93.0          & \textbf{98.1} & 0.674                     & 0.036                    \\
VT                         & RF                  & 93.3              & 92.0          & 97.7          & 0.328                     & 0.028                    \\
PCA                        & RF                  & 93.1              & 91.7          & 97.6          & 0.870                     & 0.026                    \\
AE-1L                      & RF                  & 93.2              & 91.7          & 97.7          & 1.281                     & 0.029                    \\
AE-3L                      & RF                  & 91.9              & 91.1          & 97.2          & 2.087                     & 0.028                    \\ \hline
OD                         & AdaBoost\rlap{\textsuperscript{10}}            & 90.6              & 90.6          & 96.4          & 1.488                     & 0.063                    \\
VT                         & AdaBoost            & 89.1              & 88.7          & 95.3          & 0.352                     & 0.031                    \\
PCA                        & AdaBoost            & 89.9              & 89.1          & 96.1          & 0.828                     & 0.029                    \\
AE-1L                      & AdaBoost            & 91.1              & 89.8          & 96.6          & 2.881                     & 0.048                    \\
AE-3L                      & AdaBoost            & 89.0              & 87.6          & 95.4          & 5.029                     & 0.048                    \\ \hline
OD                         & DNN-2L\rlap{\textsuperscript{11}}              & 93.0              & 93.5          & 93.0          & 162.624                   & 0.195                    \\
VT                         & DNN-2L              & 92.6              & 91.7          & 92.6          & 175.512                   & 0.225                    \\
PCA                        & DNN-2L              & 86.9              & 86.1          & 86.9          & 160.566                   & 0.259                    \\
AE-1L                      & DNN-2L              & 93.1              & 91.4          & 93.2          & 162.066                   & 0.200                    \\
AE-3L                      & DNN-2L              & 87.6              & 87.0          & 87.6          & 165.436                   & 0.215                    \\ \hline
OD                         & DNN-4L\rlap{\textsuperscript{12}}              & 93.0              & 93.0          & 94.2          & 191.264                   & 0.310                    \\
VT                         & DNN-4L              & 93.1              & 92.0          & 93.2          & 207.849                   & 0.400                    \\
PCA                        & DNN-4L              & 90.2              & 87.3          & 90.2          & 207.461                   & 0.434                    \\
AE-1L                      & DNN-4L              & 93.1              & 91.0          & 93.1          & 193.373                   & 0.333                    \\
AE-3L                      & DNN-4L              & 88.9              & 86.0          & 88.9          & 200.711                   & 0.355                    \\ \hline
OD                         & DNN-7L\rlap{\textsuperscript{13}}              & 93.0              & 92.8          & 94.0          & 254.377                   & 0.521                    \\
VT                         & DNN-7L              & 92.6              & 90.5          & 92.6          & 269.323                   & 0.664                    \\
PCA                        & DNN-7L              & 89.5              & 89.2          & 89.5          & 264.253                   & 0.717                    \\
AE-1L                      & DNN-7L              & 92.5              & 89.3          & 92.6          & 247.947                   & 0.557                    \\
AE-3L                      & DNN-7L              & 88.3              & 83.6          & 88.4          & 255.443                   & 0.598                    \\ \hline
\end{tabular}
	{\scriptsize\textsuperscript{1}Original Data}
	{\scriptsize\textsuperscript{2}Variance Threshold}
	{\scriptsize\textsuperscript{3}Principal Component Analysis}
	{\scriptsize\textsuperscript{4}AE with 1 Layer}
	{\scriptsize\textsuperscript{5}AE with 3 Layers}
	{\scriptsize\textsuperscript{6}Decision Tree}
	{\scriptsize\textsuperscript{7}k-Nearest Neighbour}
	{\scriptsize\textsuperscript{8}Support Vector Machine}
	{\scriptsize\textsuperscript{9}Random Forest}
	{\scriptsize\textsuperscript{10}Adaptive Boosting}
	{\scriptsize\textsuperscript{11}DNN with 2 Layers}
	{\scriptsize\textsuperscript{12}DNN with 4 Layers}
	{\scriptsize\textsuperscript{13}DNN with 7 Layers}
\end{table}

\section{Conclusion and Future Work}

Over the last decade, smartphones and Android OS have been growing at a tremendous pace. Such devices store a lot of personal user information which is an obvious target of cybercriminals. Research shows detection engines based on signature and heuristic methods will be unable to cope with next-generation malware.

Our extensive analysis indicates that feature engineering produces efficacious Android malware detection without significant reduction in model accuracy. We found that instead of considering all the permissions for building classification model, only $16$ permissions acquired from feature reduction (variance threshold / auto-encoder / principal component analysis) saves considerable model train and test time without significant penalty on accuracy.

Overall the baseline random forest model built with original data was able to balance both TPR and TNR to achieve highest AUC score of $98.1$. Also, Tree-based classification models like random forest and decision tree are more accurate, time-efficient and have high interpretability, which makes them suitable for real-time deployment. Also, deep neural network based models achieved a comparable accuracy but with a massive penalty of training and testing time. Furthermore, malicious applications tend to use a similar permission set which can be explained by high TPR of k-nearest neighbour based models.

We are also designing an online tool for Android malware detection that can perform real-time analysis of Android applications. In addition to the above, we are also analyzing other deep learning-based feature reduction techniques like variational autoencoder coupled with classification models built with a recurrent neural network, long short-term memory, echo state network, deep belief networks, etc. for more effective and efficient detection of Android malicious apps which may be published elsewhere.

%
%
%
\bibliographystyle{plain}
\bibliography{main}
\end{document}